\documentclass[fleqn,10pt]{wlscirep}
\usepackage[utf8]{inputenc}
\usepackage[T1]{fontenc}
\usepackage{pdfpages}
\usepackage{lineno}
\title{Carbon emissions and sustainability of launching 5G mobile networks in China}

\author[1]{Tong Li}
\author[2]{Li Yu}
\author[1]{Yibo Ma}
\author[3]{Tong Duan}
\author[1]{Wenzhen Huang}
\author[2]{Yan Zhou}
\author[1]{Depeng Jin}
\author[1]{\\Yong Li*}
\author[4]{Tao Jiang*}
\affil[1]{Beijing National Research Center for Information Science and Technology (BNRist), Department of Electronic Engineering, Tsinghua University, Beijing, China}
\affil[2]{China Mobile Research Institute, Beijing, China}
\affil[3]{National Digital Switching System Engineering and Technological Research Center, Zhengzhou, China}
\affil[4]{Research Center of 6G Mobile Communications, Huazhong University of Science and Technology, Wuhan, China.}
\affil[*]{liyong07@tsinghua.edu.cn; taojiang@hust.edu.cn}

\begin{abstract}
Since 2021, China has deployed more than 2.1 million 5G base stations to increase the network capacity and provide ubiquitous digital connectivity for mobile terminals. However, the launch of 5G networks also exacerbates the misalignment between cellular traffic and energy consumption, which reduces carbon efficiency — the amount of network traffic that can be delivered for each unit of carbon emission. In this study, we develop a large-scale data-driven framework to estimate the carbon emissions induced by mobile networks. We show that the decline in carbon efficiency leads to a carbon efficiency trap, estimated to cause additional carbon emissions of {23.82 $\pm$ 1.07} megatons in China. To mitigate the misalignment and improve energy efficiency, we propose DeepEnergy, an energy-saving method leveraging collaborative deep reinforcement learning and graph neural networks. DeepEnergy models complex collaboration among cells, making it possible to effectively coordinate the working state of tens of thousands of cells, which could help over {71\%} of Chinese provinces avoid carbon efficiency traps. In addition, applying DeepEnergy is estimated to reduce {20.90 $\pm$ 0.98} megatons of carbon emissions at the national level in 2023. We further assess the effects of adopting renewable energy and discover that the mobile network could accomplish more than 50\% of its net-zero goal by integrating DeepEnergy and solar energy systems. Our study provides insight into carbon emission mitigation in 5G network infrastructure launching in China and overworld, paving the way towards achieving sustainable development goals and future net-zero mobile networks.
\end{abstract}

\begin{document}
 \flushbottom
\maketitle

\section*{Introduction}
Connectivity has become a defining feature of the modern economy and society~\cite{manyika2011great}. 5G mobile networks play a big role by being recognized as providing high network capacity and ubiquitous digital connectivity for a massive number of terminals~\cite{dang2020should}, including but not limited to smartphones, vehicles, and sensors. 5G connectivity will usher in a new era of the digital economy, unlocking a series of innovative services, including healthcare~\cite{mwangama2020can}, autonomous vehicles~\cite{gohar2021role}, smart cities~\cite{dai2019human}, and intelligent manufacturing~\cite{taleb2019orchestrating}. According to a report from the Global Mobile Suppliers Association (GSA)~\cite{GSA2022}, more than 70 countries had launched 5G networks by June 2022. Among countries with ambitious plans to deploy 5G, China has been the global leader in commercializing 5G networks. By the end of August 2022, China has set up more than 2.1 million 5G base stations~\cite{Juan2022}, which accounts for more than 60\% of the total 5G base stations worldwide.

On the other side, the primary concern for launching 5G is the high energy consumption. Compared to previous generations of mobile networks, 5G networks have more antennas~\cite{al2022massive} and larger bandwidths~\cite{hecht2016bandwidth}, which dramatically increase the energy consumption of base stations. China Mobile's measurement report~\cite{han2020energy} indicates that the energy consumption of a 5G base station is 4.3 KWh, which is four times that of a 4G base station of 1.1 KWh. One 5G base station is estimated to produce 30 tons of carbon emissions for one year of operation~\cite{ding2022carbon}. Thus, 5G networks in China are estimated to produce over 60 megatons of carbon emissions annually at the national level. Such high energy consumption and carbon emissions would cause severe environmental problems. In order to avoid or mitigate the negative environmental impacts, we need to collect more detailed and concrete evidence from large-scale real-world data~\cite{ilieva2018social} to reveal and quantify the greenhouse impacts caused by the launch of 5G networks, and further explore a sustainable development pathway for mobile networks in China.

In this study, we delve into the sustainability of 5G mobile networks in China, aiming to find an environmentally friendly method to launch and operate 5G mobile networks. 
To quantify carbon emissions induced by mobile networks, we propose a simulation-based model considering both the mobile communication system and the power generation system and monitoring their complicated interactions. 
Specifically, we first collect a large-scale real-world dataset from Nanchang, a provincial capital in China, for modeling the energy consumption of mobile networks. The dataset includes energy consumption data (Dataset D1) of sampled base stations and traffic data (Dataset D2) of the entire network over several months. To the best of our knowledge, the dataset we collected is one of the largest in terms of data size and duration. Based on the energy consumption dataset, we build reliable energy consumption models of various types of base stations (Method M1), which estimate energy consumption with a high R-square of around 0.8 based on their traffic loads. By feeding traffic data of mobile networks into the energy consumption model, we obtain the estimated energy consumption. Next, we quantify the thermal coal consumption of power plants supporting functions of mobile networks by jointly modeling the energy consumption of mobile networks and power generation systems (Method M2). 

Based on the energy consumption and carbon emission estimation model, we discover the energy and carbon efficiency traps caused by the launch of 5G. Energy efficiency~\cite{yang2016interference, dang2020should, han2020energy}, defined as the ratio of mobile network traffic to energy consumption, is a crucial metric in the operation of mobile networks. We calculate the energy efficiency of mobile networks in Nanchang using the energy consumption model and network traffic data (Datasets D2). Our results indicate that the launch of 5G will lead to high energy consumption and a sharp decline in the energy efficiency of the entire mobile networks, which include both 4G and 5G networks: the energy efficiency in Nanchang pre-5G is {2.02 TByte/MWh}, and is estimated to drop to {1.42 TByte/MWh} after the launch of 5G (Supplementary \figurename~1).
The lower energy efficiency makes the mobile system consume extra energy to support the same amount of network traffic, forming an energy efficiency trap. 
According to our estimation model (Method M2), China is expected to consume an extra {35.02 $\pm$ 0.33} Terawatt-hour (TWh) of energy, equivalent to an additional {23.82 $\pm$ 1.07} megatons of carbon emissions. The additional carbon emissions are approximately equivalent to {80\%} of the annual carbon emissions of the power sector in France~\cite{Francestatista}. Therefore, if our societies do not take effective actions to cut down on the additional carbon emissions caused by the launch of 5G, the negative environmental impact could be catastrophic and irreversible. By exploring the link between energy efficiency and the launch of 5G, we identify that the decrease in energy efficiency is primarily due to the exacerbated misalignment between cellular traffic and energy consumption. Specifically, the misalignment occurs when a network's energy consumption is not directly proportional to its traffic load in spatial and temporal domains (see Supplementary Note 1 for details).

To mitigate the misalignment and improve energy efficiency, we propose DeepEnergy, an energy-saving method leveraging collaborative deep reinforcement learning and graph neural networks to adaptively control the working state of base stations based on their dynamic traffic loads (Method M3). When a base station's traffic load is relatively low, DeepEnergy will proactively switch the base station into a low-power operation mode to mitigate the misalignment between traffic loads and energy consumption. By implementing DeepEnergy on Nanchang's mobile networks, the misalignment factors\footnote{A metric to measure the severity of misalignment between traffic loads and energy consumption, which is in the interval $[0,1]$. The misalignment is more severe as the factor increases. (See Supplementary Note 1 for details)} are estimated to reduce from 0.56 to 0.28 on average. As a result, the energy efficiency and carbon efficiency will reach {2.83 TByte/MWh and 4.16 TByte/tCO$_2$}, respectively, nearly twice as much as the case without employing DeepEnergy. It is estimated that DeepEnergy would successfully help over {71\%} of Chinese provinces to avoid energy efficiency traps and additional carbon emissions. In addition, applying DeepEnergy would reduce {20.90 $\pm$ 0.98} megatons of carbon emissions at the national level in China in 2023. By further integrating DeepEnergy and solar energy systems, the mobile network is estimated to accomplish over 50\% of its net-zero target. Our study paves the way for achieving sustainable development goals and future net-zero mobile networks by providing insight into carbon emission mitigation in 5G network infrastructure operations.

\section*{Results}

\subsection*{Launching 5G Leads to Carbon Efficiency Trap}

\begin{figure*}[ht]
\centering
\includegraphics[width=1\textwidth]{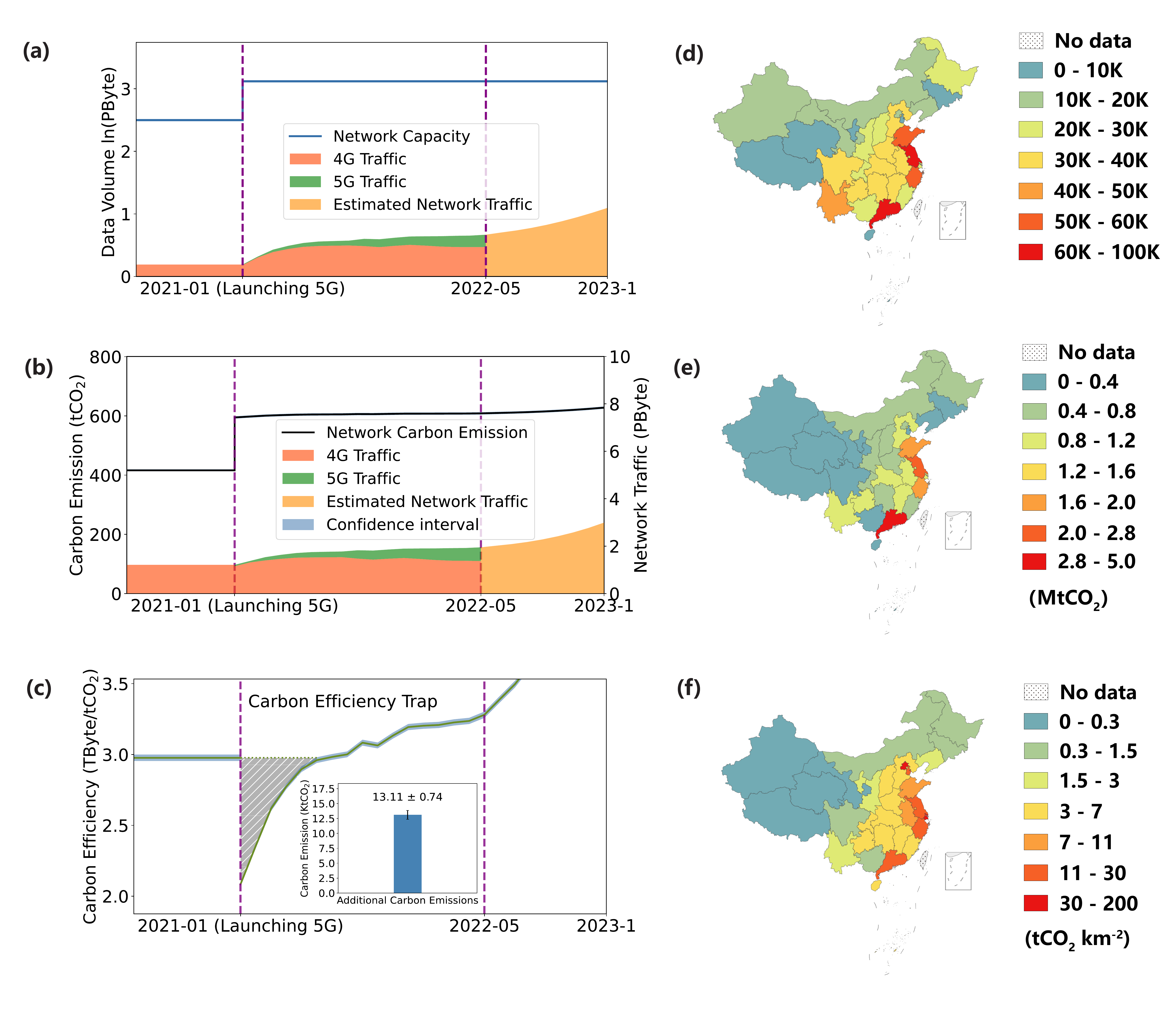}
\caption{\textbf{Analysis of the carbon efficiency after the launch of 5G networks}. (a). Launching 5G leads to a substantial increase in daily network capacities in Nanchang. (b) The operation of newly launched 5G base stations has led to a sharp increase in carbon emissions in Nanchang. (c) Launching 5G leads to the appearance of a carbon efficiency trap and causes additional carbon emission of {13.11 $\pm$ 0.74 KtCO$_2$} in Nanchang. (d). The number of 5G base stations across different provinces in China. (e). The number of additional carbon emissions across different provinces in China. (f). The number of additional carbon emissions per unit space across different provinces in China.}
\label{fig:carbontrap}
\end{figure*}

We quantify and estimate the carbon emissions of mobile network operations in Nanchang from 2020 to 2023 (Method M2.1). As illustrated in \figurename~\ref{fig:carbontrap} (a), the launch of 5G results in an increase in daily network capacities from 12 PByte (12,264 4G base stations) to 22 PByte (12,264 4G base stations and 2,159 5G base stations). The operation of newly launched 5G base stations has led to a sharp increase in energy consumption and a decline in energy efficiency (Supplementary \figurename 1). Correspondingly, there has been a dramatic increase in daily carbon emissions of {178} tons after launching 5G (\figurename~\ref{fig:carbontrap} (b)). {Carbon efficiency, the amount of network traffic that can be delivered for one unit of carbon emissions, decreases from 2.98 TByte/tCO$_2$ to 2.08 TByte/tCO$_2$ (\figurename~\ref{fig:carbontrap} (c)). 
After launching the 5G network, there is a large and rapid increase in carbon emissions. Nevertheless, the traffic load does not grow in the same trend, exacerbating the misalignment between cellular network traffic and energy consumption, which is the critical reason for lowering the carbon efficiency of the mobile network.}
Although the carbon efficiency does increase over time as mobile users consume more network traffic, Nanchang is estimated to take more than {six} months to return to its pre-5G level of carbon efficiency. This decrease in carbon efficiency and subsequent recovery process is referred to as the carbon efficiency trap, as depicted by the grey shadow area in \figurename~\ref{fig:carbontrap} (c). In the carbon efficiency trap, the mobile network will produce additional carbon emissions to carry the same network traffic due to the reduction in carbon efficiency. In Nanchang, for example, the efficiency trap is estimated to cause additional {13.11 $\pm$ 0.74} Kilotons of carbon dioxide. 

The energy consumption of a mobile network depends on the number of base stations (Supplementary \figurename~2). We next generalize the simulation results from Nanchang to all provinces in China using the Monte Carlo method by considering the number of 4G and 5G base stations in each province (Method M2.2). The uncertainty in carbon emissions caused by mobile networks in each province is approximate -6\% and +6\% at the 95\% confidence level (see Supplementary Tables 7 and 16 for details).
According to the estimation results, the eastern provincial regions have a higher number of 5G base stations and generate more additional carbon emissions (\figurename s~\ref{fig:carbontrap} (d) and~\ref{fig:carbontrap} (e)). Guangdong, Jiangsu, and Zhejiang are the top three contributors, each accounting for {0.13, 0.11, and 0.08} of the national carbon emissions. In contrast, the western provincial regions, such as Qinghai and Tibet, have fewer 5G base stations and produce fewer additional carbon emissions. 
In addition, compared to the less developed western areas, the wealthier eastern provincial regions have higher additional carbon emissions per unit area (\figurename~\ref{fig:carbontrap} (f)). Tibet has the lowest additional carbon emissions per unit area of {0.0155 tCO$_2$/km$^2$}, whereas Shanghai's additional carbon emissions per unit area are {roughly 10,811 times higher (167.58 tCO$_2$/km$^2$)}. 
Notably, the amount of carbon dioxide each square kilometer of the forest is capable of absorbing approximately 500 tons per year~\cite{birdsey1992carbon}. In other words, the additional carbon emissions caused by the launch of 5G in Shanghai would require a forest, which is the same size as Shanghai city, to take at least {four months} to absorb.
Therefore, based on the intensity and density of additional carbon emissions, China, especially the eastern provincial regions, may face severe environmental issues which would cause irreversible damage during the launch of 5G. We urgently need to dig into the root causes and find a solution to sustainably launch and operate 5G networks.

\subsection*{Cellular Traffic and Energy Consumption Show Huge Spatio-temporal Misalignment}

\begin{figure*}[ht]
\centering
\includegraphics[width=0.88\textwidth]{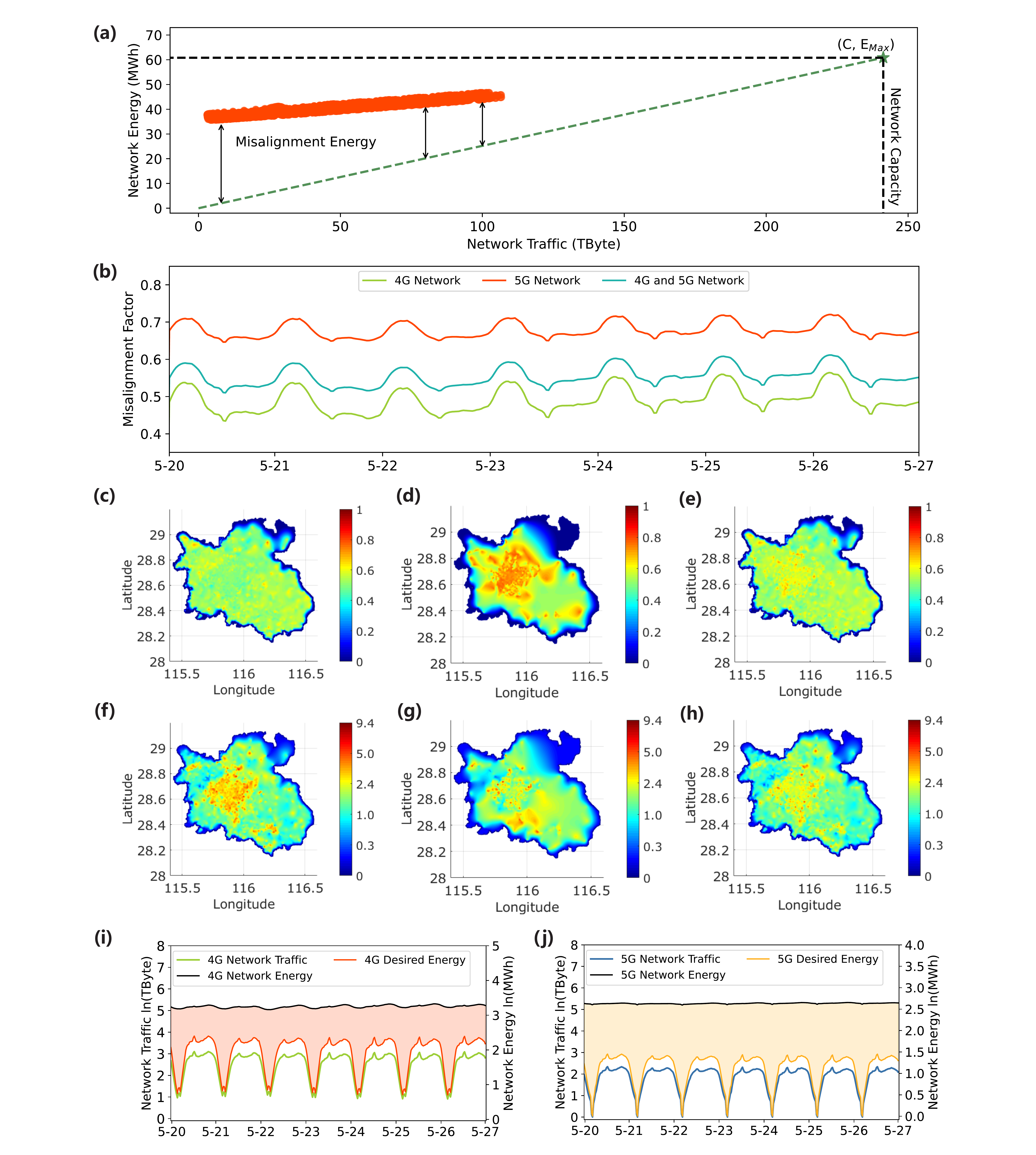}
\caption{\textbf{Cellular traffic and energy consumption show huge spatio-temporal misalignment}. (a). The energy-traffic curve for mobile networks is based on the network performance data collected in Nanchang. Network energy is not proportional to its carried traffic load, showing the misalignment between traffic and energy. (b). One-week misalignment factor pattern of mobile networks. The launch of 5G networks causes the mobile network's misalignment factor to rise. (c). Spatial distribution of misalignment factors of 4G networks. (d). Spatial distribution of misalignment factors of 5G networks. (e). Spatial distribution of misalignment factors of the entire mobile networks. (f). Spatial distribution of energy efficiency of 4G networks. (g). Spatial distribution of energy efficiency of 5G networks. (h). Spatial distribution of energy efficiency of the entire mobile networks. Regions with large misalignment factors exhibit low energy efficiency and vice versa. (i). one-week traffic and energy pattern of the 4G network. (j). one-week traffic and energy pattern of the 5G network. Addressing the spatio-temporal misalignment between cellular traffic and energy consumption in mobile networks could significantly reduce energy consumption.}
\label{fig:Misalignment}
\end{figure*}

To uncover the causes of the carbon efficiency trap, we investigate the spatio-temporal misalignment between cellular traffic and energy consumption, which refers to the fact that the energy consumption of a mobile network is not directly proportional to the traffic load. 
Taking the network traffic data of Nanchang as an example, \figurename~\ref{fig:Misalignment} (a) shows how the energy consumed by the mobile network varies with the aggregated traffic load. 
Even with light traffic of 2.5 TBytes, i.e., 1\% of the network capacity, the energy consumption is relatively high at 25.6 MWh, roughly 42\% of the maximum energy consumption. 
A mobile network reaches its maximum energy consumption, denoted by $E_{Max}$, when the carried traffic reaches its available capacity $C$.
In contrast, the desired energy consumption~\cite{peng2011traffic}, which is proportional to the traffic load (the green dashed line in \figurename~\ref{fig:Misalignment} (a)), is much less, at only about 0.6 MWh for light traffic of 2.5 TBytes.
To measure the misalignment between traffic load $L$ and energy consumption $E$, we define a metric called the misalignment factor, denoted by $M=\tilde{E} - \tilde{L}$, where $\tilde{E} = E/E_{Max}$ represents the normalized energy consumption and $\tilde{L} = {L}/{C}$ represents the normalized traffic load. The misalignment factor is in the interval $[0,1]$ to measure the discrepancy between normalized energy consumption and traffic load (see Supplementary Note 1 for details). 
It is more severe as the factor increases. 
When $M=0$, the energy consumption is directly proportional to the traffic load, and there is no misalignment. 

The misalignment factor between traffic load and energy consumption changes over time and space. 
As illustrated in \figurename~\ref{fig:Misalignment} (b), the misalignment factor shows an apparent daily pattern, with higher values at night and lower values during the day. Specifically, the misalignment factor of 5G networks (0.68) is much higher than that of 4G networks (0.49), increasing the misalignment factor of the entire mobile network to 0.56 on average. 
Regarding spatial distribution, the misalignment factors of 5G networks are higher in the city center where the 5G base stations are more concentrated (\figurename~\ref{fig:Misalignment} (d)). The misalignment factors of 4G networks are generally uniformly distributed with low values (\figurename~\ref{fig:Misalignment} (c)). Thanks to years of optimization of 4G base station deployments, the energy consumption of 4G base stations has been having a similar spatial distribution with the traffic loads (Supplementary \figurename~3). In contrast, due to the insufficient traffic profiles of 5G applications, the deployment of 5G base stations is not yet optimized, which leads to extremely high misalignment factors, often exceeding 0.8, in most regions.

{Given a mobile network's traffic load $L$, the energy efficiency, denoted by ${\eta}^{Energy}(L)$, can be expressed as ${\eta}^{Energy}(L) = {{\eta}^{Energy}_{Desired}} / {(1 + M/\tilde{L})},$ where ${\eta}^{Energy}_{Desired} = C/E_{Max}$ denotes the desired energy efficiency of that mobile network (see Supplementary Note 1 for details). Thus, the energy efficiency is affected by three key factors: $\tilde{L}$, $\eta_{Desired}^{Energy}$, and $M$. Specifically, $\eta_{Desired}^{Energy}$ depends on the technologies used in the mobile network. Thanks to the advanced technologies of 5G, such as massive MIMO and subframe silence~\cite{rostami2019pre},~\cite{rostami2019pre}, the desired energy efficiency of 5G is 6.29 TByte/MWh, which is twice that of 4G (3.01 TByte/MWh). However, the rise in the misalignment factor reduces energy efficiency.} As illustrated in \figurename~\ref{fig:Misalignment} (f)-(h), regions with large misalignment factors exhibit low energy efficiency and vice versa. The average energy efficiency of the city center falls from 2.98 TByte/MWh (4G networks) to 1.94 TByte/MWh (covering both 4G and 5G networks), corresponding to the increase in the misalignment factor. In summary, launching 5G networks increases the misalignment between traffic load and energy consumption, and lowering the energy efficiency of the entire mobile networks.

\figurename s~\ref{fig:Misalignment} (i) and (j) depict one-week patterns of 4G and 5G networks for traffic loads, energy consumption in the real world, and the desired energy consumption without misalignment, respectively.
The traffic loads of both 4G and 5G exhibit a diurnal rhythm. Whereas, the network energy consumption remains almost constant throughout the day. Such findings are consistent with \figurename~\ref{fig:Misalignment} (a). 
Also, there is a vast gap between the current energy consumption with huge misalignment and the desired energy consumption without misalignment (colored areas in \figurename~\ref{fig:Misalignment} (i)), indicating a remarkable potential to reduce energy consumption and increase energy efficiency by addressing the spatio-temporal misalignment between cellular traffic and energy consumption in mobile networks.

\subsection*{Energy-Saving Methods Mitigate Efficiency Trap and Misalignment}

The main factor behind the misalignment between traffic and energy is that the energy consumption of cooling devices and fixed radio transmission overhead at the base station are unaffected by traffic load (see Supplementary Note 2 for details). One solution to mitigate the misalignment is to switch the base station into low-power operation mode by proactively turning off some components, such as the power amplifier and cooling devices, when the traffic load is relatively low.
This low-power operation mode of the base station is known as the sleep mode~\cite{wu2015energy}. Notably, a base station in sleep mode cannot provide service. Therefore, other active compensation base stations nearby must handle mobile users who were previously served by the base station in sleep mode (see Supplementary Note 4 for details). 

We propose an artificial intelligence (AI) empowered energy-saving method, DeepEnergy, which is based on collaborative deep reinforcement learning to control the working state of cells\footnote{A cell refers to a carrier on a sector of a base station. A base station generally has multiple cells.} adaptively (Method M3). 
Specifically, DeepEnergy models each cell as an intelligent agent that self-decides its working state based on traffic loads. DeepEnergy aims to learn an action-value network for cells, which would help quantify how much energy could be saved from each cell's working state decision. 
Since the mobile network is enormous and cells may interact with each other, DeepEnergy adopts a collaborative grid-based learning strategy to learn the action-value network. DeepEnergy divides the region into small grids based on the compensation relationships of cells. The cells in each grid are equivalent in service capability, and thus can act as collaborative compensation cells for each other. Moreover, DeepEnergy models the attribution relationships between cells and base stations. As a result, DeepEnergy considers both intra-grid and intra-base station collaboration among cells, making it possible to effectively coordinate the working state of tens of thousands of cells.

\begin{figure*}[ht]
\vspace{-20mm}
\centering
\includegraphics[width=1\textwidth]{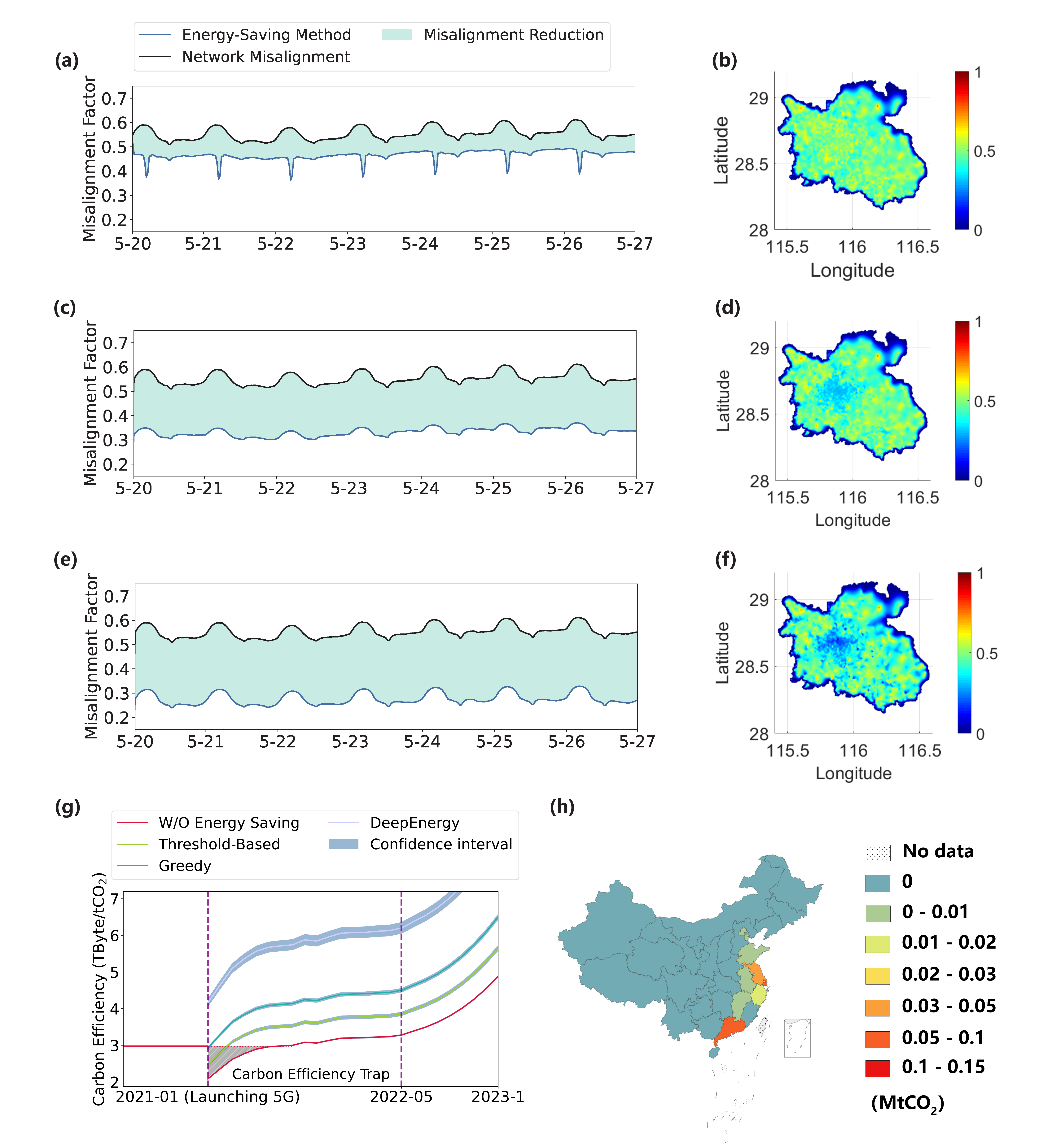}
\vspace{-3mm}
\caption{\textbf{Performance analysis of energy-saving methods}. (a). Temporal distribution of misalignment factors of the entire mobile networks in Nanchang using the threshold-based energy-saving method. (b). Spatial distribution of misalignment factor of the entire mobile networks in Nanchang using the threshold-based energy-saving method. (c). Temporal distribution of misalignment factors of the entire mobile networks in Nanchang using the greedy energy-saving method. (d). Spatial distribution of misalignment factors of the entire mobile networks in Nanchang using the greedy energy-saving method. (e). Temporal distribution of misalignment factors of the entire mobile networks in Nanchang using DeepEnergy. (f). Spatial distribution of misalignment factors of the entire mobile networks in Nanchang using DeepEnergy. (g). The energy-saving methods have effectively improved carbon efficiency, and the DeepEnergy method can effectively avoid the carbon efficiency trap caused by launching the 5G network. (h). The number of additional carbon emissions across different provinces in China after using DeepEnergy. Seventeen provinces have zero additional carbon emissions, and other provinces have seen significant reductions in additional carbon emissions.}
\label{fig:SleepMode}
\vspace{-2mm}
\end{figure*}

We compare DeepEnergy with two classical energy-saving methods: threshold-based method~\cite{yu2014dual} and greedy method~\cite{peng2011traffic} (see Supplementary Note 4 for details).
Overall, DeepEnergy outperforms others on the entire mobile network, consisting of 4G and 5G, reducing the misalignment factors from 0.56 to 0.28 on average (\figurename~\ref{fig:SleepMode} (e)). 
The threshold-based method performs the worst because there is no inter-cell collaboration, and cells independently decide whether to enter sleep mode based on a predefined traffic load threshold.
As a result, the threshold-based method can only barely work at night when base station traffic is typically low (\figurename s~\ref{fig:SleepMode} (a)). Alternatively, the greedy method considers intra-grid collaboration and has a comparable performance with DeepEnergy (\figurename~\ref{fig:SleepMode} (c)). Regarding the spatial distribution, DeepEnergy performs noticeably better in the city center with a misalignment factor of 0.22, which is much lower than that of the greedy method (0.53) and of the threshold-based method (0.41) (\figurename s~\ref{fig:SleepMode} (b), (d), and (f)). This is because the city center's base station density is comparatively high. DeepEnergy considers both intra-grid and intra-base station collaborations of cells, making it more effective in dealing with regions with dense base station deployments and complex interactions between cells. Considering that the average density of 5G base stations is expected to be three times higher than that of 4G in the future~\cite{han2020energy}, DeepEnergy has greater potential for energy savings in future mobile networks with super base station densities.

Accredit to the reduction in misalignment factors, DeepEnergy significantly improves the energy efficiency and carbon efficiency of mobile networks, reaching {2.83 TByte/MWh and 4.16 TByte/tCO$_2$}, almost doubled in comparison to the case without energy-saving (\figurename~\ref{fig:SleepMode} (g)). 
We next generalize the results from Nanchang to all provinces in China (Method M2.2). Our results show that energy-saving methods can reduce the additional carbon emissions caused by launching 5G in China {from 23.82 $\pm$ 1.07 megatons to 11.34 $\pm$ 0.46 megatons (threshold), to 4.24 $\pm$ 0.17 megatons (greedy), and to 0.18 $\pm$ 0.01 megatons (DeepEnergy)}. Specifically, DeepEnergy can help more than {71\%} of provinces in China successfully avoid the energy efficiency trap without producing additional carbon emissions (\figurename~\ref{fig:SleepMode} (h)). In addition, applying DeepEnergy would continue to reduce carbon emissions every year, which is estimated to reduce 25.12 out of 50.52 megatons (2021), 24.65 out of 51.03 megatons (2022), and 20.90 out of 55.42 megatons (2023) at the national level (see Supplementary Tables 9-11 and 16 for details).

\subsection*{Renewable Energy Helps Achieve Net-zero Mobile Networks}

\begin{figure*}[ht]
\vspace{-2mm}
\centering
\includegraphics[width=0.98\textwidth]{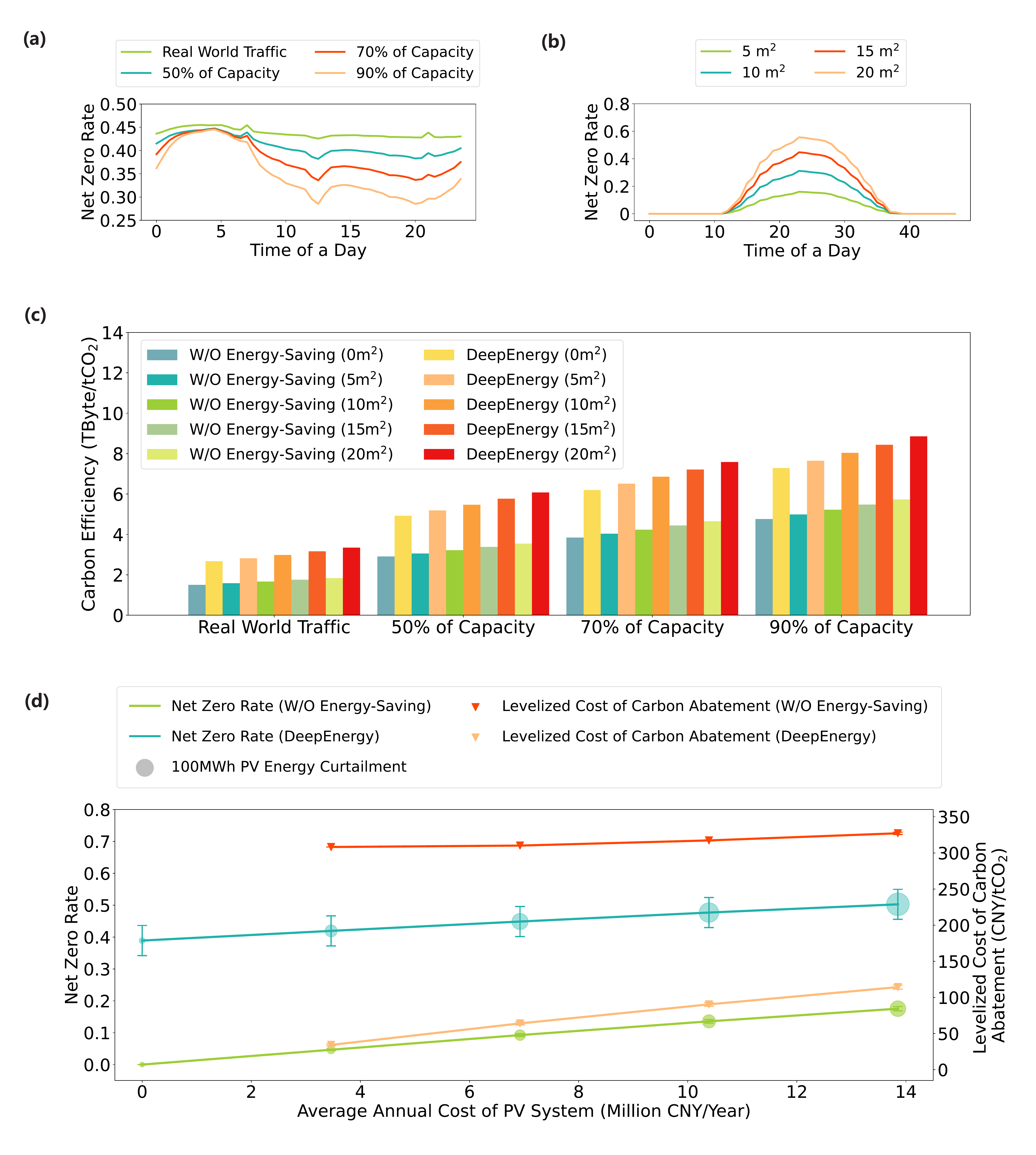}
\caption{\textbf{Renewable energy helps achieve net-zero mobile networks}. (a). Net-zero rate using DeepEnergy in a day under different traffic loads. {50\%, 70\%, 90\% of capacity refer to future counterfactual cases of change in traffic volume. For example, 50\% of capacity means that the maximum traffic volume of each cell in a day reaches 50\% or above of its capacity.}
The performance of DeepEnergy at night is superior to that of the daytime due to diurnal rhythms. (b). Net-zero rate of a PV system in a day under different panel sizes. PV power generation systems can operate in the daytime and deliver clean energy to the mobile network. (c). The carbon efficiency of the DeepEnergy versus the without (W/O) energy-saving scenario for various PV systems. Combining DeepEnergy and the PV system can significantly improve the carbon efficiency of a mobile network. (d). Net-zero rates and levelized cost of carbon abatement under different economic investments in the PV system. The radius of each bubble represents the energy curtailment for the solar energy system.}
\label{fig:PVSystem}
\vspace{-2mm}
\end{figure*}

Due to diurnal rhythms, DeepEnergy performs better at night than in the daytime (\figurename~\ref{fig:PVSystem}~(a)). Also, with the increase in traffic, the carbon reduction ratio will decrease significantly during the day, dropping below 0.3 when the traffic load reaches 90\% of network capacity. Thus, DeepEnergy will have limitations in terms of energy-saving during the daytime. Fortunately, the photovoltaic (PV) power system can operate during the day and provide renewable energy to the mobile network, further reducing carbon emissions in the daytime (\figurename~\ref{fig:PVSystem}~(b)). By deploying PV panels on base stations~\cite{chamola2016solar}, DeepEnergy and the PV power system can work together to reduce carbon emissions both day and night.

We next assess the carbon efficiency of the mobile network in Nanchang when we jointly operate DeepEnergy and the PV system through our proposed simulation-based system. Specifically, we locate PV panels at each base station and use PVWatts~\cite{freeman2018system, waldman2019solar} to estimate the potential electric generation of the PV system (Method M4). \figurename~\ref{fig:PVSystem}~(c) depicts the carbon efficiency of PV systems with various sizes of PV panels for DeepEnergy and the case without energy-saving. The carbon efficiency of PV system using DeepEnergy is much higher than the case without energy-saving. The overall carbon efficiency increases as cellular traffic boosts. The gap in carbon efficiency between DeepEnergy and the case without energy-saving widens {from 1.50 TByte/tCO$_2$ to 3.12 TByte/tCO$_2$} when the traffic load reaches 90\% of network capacity. Therefore, combining DeepEnergy and the PV system can significantly improve the carbon efficiency of a mobile network.

The establishment of PV systems must take into account the economic investment~\cite{peng2018managing}. We next estimate the costs of construction and maintenance of PV systems and reveal how system performance changes with various economic investments. As illustrated in \figurename~\ref{fig:PVSystem} (d), as investment increases, the mobile network becomes closer to the goal of net-zero. The mobile network is estimated to achieve over 50\% of its net-zero targets with DeepEnergy and a PV system when the annual investment contributed to the PV system is 13.85 million CNY. However, as the size of the PV system increases, the energy curtailment will become severe~\cite{Wang:20, bogdanov2019radical}, lowering the cost efficiency due to marginal utility. We find that applying solar energy alone to reduce carbon emissions is not enough to achieve cost-effectiveness, with levelized cost of carbon abatement (LCCA) of {308.18 CNY/tCO$_2$}. The LCCA can be reduced to about one ninth when we integrate DeepEnergy, which is {34.29 CNY/tCO$_2$}. As a result, DeepEnergy can help solar energy become an affordable path toward net-zero mobile networks in terms of cost efficiency.

\section*{Discussion and Implications}

Our study reveals that the launch of 5G will lower the energy efficiency of the entire mobile network due to the high misalignment between traffic loads and energy consumption. The decline in energy efficiency will result in significant additional energy consumption and additional carbon emissions, causing irreparable harm to the environment. Therefore, we are facing urgent needs for an environmentally friendly launch and operation strategy and must alter how we currently launch and operate 5G mobile networks. 
In this paper, we propose an energy-saving method, DeepEnergy, to proactively control base stations' working states based on their dynamic traffic loads. DeepEnergy can mitigate the misalignment between traffic and energy in mobile networks, remarkably improving energy efficiency. As a result, in addition to assisting cities in avoiding energy efficiency traps, DeepEnergy continues to reduce energy consumption over time. For instance, it is estimated that from January 2021 to January 2023, the energy efficiency gain of Nanchang introduced by DeepEnergy will rise from {1.41 TByte/MWh} to {3.73 TByte/MWh}, and DeepEnergy will reduce {30.74 TWh} of energy consumption in 2023.
Furthermore, by integrating DeepEnergy and solar energy, the mobile network is estimated to accomplish over 50\% of its net-zero target. Consequently, DeepEnergy can assist in the launch of 5G and make the mobile network infrastructure update process more environmentally friendly. {However, there are several potential obstacles that might be encountered when implementing DeepEnergy. Firstly, DeepEnergy is a machine learning model requiring significant data to train and optimize. The availability and quality of the data could be a potential limitation. Secondly, implementing DeepEnergy at scale would require computational resources and technical expertise, which could be hard to achieve for some organizations. Finally, operators may emphasize encouraging the usage of 5G networks and be reluctant to set 5G base stations to sleep, which may cause the operators to forgo DeepEnergy for commercial reasons.}

As an initial attempt, our study still has some limitations. Our analysis is conducted based on real-world datasets collected from China Mobile. 
The carbon emissions of mobile networks are underestimated because China Mobile only accounts for a portion of the market share (over 58\%). 
We generalize the simulation results from Nanchang to all the provinces in China, which would cause uncertainty.
Renewable energy resources in different provinces in China are expected to be uneven. Many regions in western China have abundant indigenous renewable energy resources. The uncertainty resulting from differences in the distribution of energy resources will affect the path toward net-zero goals. In China, the deployment of new 5G base stations is still ongoing. Our study only examined the first wave of 5G base station launches, which may lead to an underestimation of carbon emissions. 

Sustainable development goals (SDGs) are urgent call for actions for all industries. Our study quantifies the environmental impacts of launching 5G mobile networks and proposes a practical approach to achieve the SDGs in mobile networks by integrating energy-saving methods and solar energy systems. 
Our study offers a firm foundation for further research into the environmental deployment and operation of 5G networks. Additionally, as newer AI architectures and more powerful reinforcement learning models continue emerging, our study opens the door for developing more sophisticated systems for carbon emission reduction in mobile networks. The cost of deploying AI algorithms on base stations is much lower than the investment in renewable energy equipment, making it more affordable for developing countries to launch 5G environmentally friendly. Consequently, AI technologies not only help achieve the SDGs in mobile networks accredit to their superior performance, but also mitigate digital inequality among developed and developing countries because of low deployment costs. 

\section*{Datasets}
In this paper, we simulate the carbon emissions from mobile networks in Nanchang and then extrapolate the results to all provinces in China. The following data are collected to assess the investigation: energy consumption data of base stations, network traffic data in Nanchang, and the number of base stations and mobile users in each province.

\subsection*{D1. Energy Consumption Data of Base Stations}
Energy consumption data are essential for building the energy consumption model of base stations. 
We collected the data from 300 4G base stations and 266 5G base stations in real-world mobile networks over one week in May 2022. The energy consumption records of base stations are collected every half hour. Each record includes the base station ID, timestamp, physical resource block (PRB) usage ratio, traffic volume, the number of users served, transmit power, baseband unit (BBU) power, and remote radio unit (RRU)~\footnote{In 5G networks, this part is called the active antenna processing unit (AAU). In this paper, we uniformly refer to the RRU for both 4G and 5G.} power (see Supplementary Table 1 for details). The energy consumption dataset covers four types of 5G base stations and three types of 4G base stations, each with a different maximum transmit power setting (see Supplementary Note 2 for details).

\subsection*{D2. Network Traffic Data in Nanchang}
Network traffic load is a crucial factor that determines the energy consumption of mobile networks. The network traffic data was also collected from China Mobile. We carried out a city-level measurement in Nanchang and collected fine-grained records on the network traffic of all 4G and 5G base stations for one week in May 2022. The network traffic data covers 12,264 4G base stations and 2,159 5G base stations. Each network traffic record is collected every half hour and contains the base station ID, timestamp, PRB usage ratio, traffic volume, and the number of users served (see Supplementary Table 1 for details). The fine-grained dataset is to reveal the mobile network's daily energy consumption patterns. We also conducted a long-term monthly measurement in Nanchang and collected its monthly network traffic for 4G and 5G networks over 17 months, from January 2021 to May 2022 (see Supplementary Table 2 for details). The monthly dataset is used to examine the long-term trends in energy consumption and carbon emissions.  

\subsection*{D3. Number of Base Stations and Mobile Users in Each Province}
We gathered the number of 4G and 5G base stations in each province in China to generalize the results from Nanchang to all provinces because the quantity of energy consumption and carbon emissions of mobile networks are directly related to the number of base stations (Supplementary~\figurename~2). Also, we gathered the number of mobile users in each province to estimate its network traffic since the average traffic per user across regions is comparable. The Ministry of Industry and Information Technology of the People's Republic of China~\cite{miitgov} releases monthly data on the number of base stations and mobile users in each province (see Supplementary Tables 3 and 4 for details).

\section*{Methods}

In this paper, we first build a reliable data-driven energy consumption model of base stations, which estimates energy consumption based on base stations' traffic loads. Next, we develop a carbon emission estimation model consisting of two parts: carbon emission estimation in Nanchang and carbon emission estimation across provinces. The estimation model considers both the mobile communication system and the power generation system and monitors the complicated interactions between these two systems. We next propose an energy-saving method called DeepEnergy, which uses a collaborative multi-agent deep reinforcement learning method that learns to optimize the working states of cells dynamically. Lastly, we analyze solar energy systems to discover the net-zero path of the mobile network.

\subsection*{M1. Energy Consumption Model of Base Stations} 
Base stations are fundamental elements of mobile networks and are their principal energy consumers~\cite{han2020energy}. A base station generally consists of a communication subsystem and a supporting subsystem (see Supplementary \figurename~7 for details). The Remote Radio Unit (RRU) and Baseband Unit (BBU), major components of the communication subsystem, are responsible for transceiving radio signals and processing baseband signals, respectively. A base station may have several RRUs and BBUs. The cooling and other auxiliary devices are part of the supporting subsystem. The cooling equipment, like air conditioning, is used to keep the base station at a proper operating temperature. The total power consumption $P_{BS}$ of a base station can be calculated as follows,
\begin{linenomath*}
\begin{equation}
P_{BS} = P_{tx} + P_{cooling},
\end{equation}
\end{linenomath*}
where $P_{tx}$ denotes the power consumption of the communication subsystem and $P_{cooling}$ denotes the power consumed by cooling equipment to maintain an appropriate operating temperature. In this paper, we only model the cooling devices for the supporting subsystem because they account for over 90\% of the supporting system’s power consumption~\cite{arnold2010power}.

The power consumption of the communication subsystem $P_{tx}$ is dominated by two components,
\begin{linenomath*}
\begin{equation}
\label{eq:tx}
P_{tx} = P_{BBU} + P_{RRU},
\end{equation}
\end{linenomath*}
where $P_{RRU}$ varies according to the amount of base station traffic. When the traffic load is heavy, RRU has to consume more power to support more active physical resource blocks (PRBs)\footnote{A physical resource block is the smallest unit of resources allocated to a mobile user for carrying network traffic.}. On the other hand, baseband processing is handled by BBUs. $P_{BBU}$ is proportional to the number of BBUs equipped at base stations. No matter how many PRBs are active, their power consumption remains constant and does not change in response to base station traffic. {By analyzing real-world data, we find that the BBU power of a base station is related to the number of cells it has (Supplementary \figurename 8) because a base station with more cells generally has more BBUs. There are three types of base stations (BSes) in terms of their number of cells and network types: 4G BS with three cells, 5G BS with three cells, and 5G BS with six cells. Since the BBU power of a particular BS type is mainly distributed in a small interval ($\pm8.88\%$), we use the average value to approximate it (Supplementary Table 20). Specifically, for a 4G BS with three cells, its BBU power is simulated as 89.3771 W. For a 5G BS with three cells, its BBU power is simulated as 305.0409 W. For a 5G BS with six cells, its BBU power is simulated as 499.6484 W.}

According to the energy consumption data of base stations, $P_{RRU}$ varies from 200W to 1200W and shows a linear relation with transmit power $P_{trans}$ (Supplementary \figurename s~9 and 11). 
\begin{linenomath*}
\begin{equation}
\label{eq:RRU}
P_{RRU} = \alpha \cdot P_{trans} + \gamma,
\end{equation}
\end{linenomath*}
where $\alpha$ and $\gamma$ denote slope and offset, respectively. In practice, $\alpha$ depends on the power amplifier efficiency of base stations, and $\gamma$ depends on fixed circuit power~\cite{CL2014Toward}. The linear regression model can approximate the $P_{RRU}$ of various base stations well, with an R-squared of over 0.75 for 5G base stations (Supplementary \figurename~10) and over 0.93 for 4G base stations (Supplementary \figurename~11). Different base station types have specific coefficients in the linear regression model, as depicted in Supplementary Tables 21 and 22. The offset $\gamma$ of 5G base stations is typically 2 to 7 times greater than that of 4G base stations, reflecting the extremely high fixed circuit power of 5G base stations. 

The transmit power $P_{trans}$ ranges from 0W to 300W. We find that transmit power varies with the PRB usage ratio, and a linear regression model can approximate this relationship (Supplementary \figurename s~12 and 14). 
\begin{linenomath*}
\begin{equation}
\label{eq:trans}
P_{trans} =  \beta \cdot r_{_{PRB}} + \sigma,
\end{equation}
\end{linenomath*}
where $r_{_{PRB}}$ denotes the PRB usage ratio, $\beta$ and $\sigma$ denote slope and offset, respectively. As depicted in Supplementary Tables~21 and 22, base stations of different types also have different coefficients. The linear regression model can model transmit power with an R-squared of over 0.96 for 5G base stations and 1 for 4G base stations. Also, since 5G base stations adopt more advanced physical technologies, e.g., subframe shutdown~\cite{Huang2020Prospect}, the offset $\sigma$ in \eqref{eq:trans} equals 0. By substituting \eqref{eq:trans} into \eqref{eq:RRU}, we obtain the relationship between RRU power ($P_{RRU}$) and PRB usage ratio ($r_{_{PRB}}$). Given a base station's traffic load $L$, $r_{_{PRB}}$ can be approximated as $L/C$, where $C$ denotes the base station's capacity. 

Alternatively, when a base station is in sleep mode, its RRU power becomes independent of its traffic load and principally distributed into a small interval (Supplementary \figurename~15). Therefore, we use the average value to approximate the RRU power in sleep mode for different base station types (Supplementary Table 23). The RRU power in the sleep mode of 5G base stations ranges from 69.43W to 90.56W, which is significantly lower than that of 4G base stations, which range from 119.03W to 133.90W. Thanks to the advanced physical energy-saving technology utilized by 5G~\cite{Lopez-Perez2022}.

The power consumption of the cooling devices is simulated using EnergyPlus~\cite{crawley2001energyplus}, a widely used program developed by Lawrence Berkeley National Laboratory (LBNL). EnergyPlus can simulate the power consumption of the cooling system every half hour, given the outdoor air temperature, the indoor proper operating temperature, the power generated by the communication subsystem, and the room size of a base station. 
Specifically, the outdoor air temperature data can be found in the World Meteorological Organization's weather dataset\footnote{\href{https://worldweather.wmo.int/en/home.html}{https://worldweather.wmo.int/en/home.html}}. The communication subsystem is modeled as an electrical heat source, and its power consumption is estimated according to \eqref{eq:tx}. According to the ``Technical Standard for Mobile Communication Infrastructure of Construction'' recognized by the Chinese government, we set the room size of a base station as 20$m^2$ and 20$^{\circ}$C as the indoor proper operating temperature. 

\subsection*{M2. Carbon Emission Estimation Model}
We develop a carbon emission estimation model consisting of two parts: carbon emission estimation in Nanchang and carbon emission estimation across provinces. The first part quantifies the thermal coal consumption of power plants in Nanchang by jointly modeling the mobile network and the power generation system and monitoring the complicated interactions between these two systems. The second part aims to expand the results of Nanchang to all of China's provinces.

\subsubsection*{M2.1. Carbon Emission Estimation in Nanchang} 

\textbf{Electricity Dispatch Model.} Electricity dispatching, which selects available generating units in a region to meet the power loads, is typically used to schedule the amount of power generated by each plant. In this paper, a day-ahead unit commitment model is formulated to quantify the thermal coal consumption of power plants, which is expressed as a linear programming optimization problem: 
\begin{linenomath*}
\begin{equation} 
\label{eq:Dispatch1}
\min_{X} \sum_{i\in \Phi} \sum_{t\in T} [c_i^{power}x_{i}^{power}(t)+c_i^{up}x_{i}^{up}(t)+c_i^{down}x_{i}^{down}(t)],
\end{equation}
\end{linenomath*}
subject to:
\noindent
\begin{linenomath*}
\begin{equation} 
\label{eq:Dispatch2}
	\sum_{i\in \Phi} x_{i}^{power}(t)+P_{outside}(t) = P_{load}(t)/(1-r_{loss}), \forall t\in T,
\end{equation}
\end{linenomath*}
\begin{linenomath*}
\begin{equation} 
\label{eq:Dispatch3}
	P_{i}^{min}x_{i}^{on/off}(t) \leqslant x_{i}^{power}(t) \leqslant P_{i}^{max}x_{i}^{on/off}(t), \forall i\in \Phi, \forall t\in T,
\end{equation}
\end{linenomath*}
\begin{linenomath*}
\begin{equation} 
\label{eq:Dispatch4}
	\sum_{i\in \Phi} P_{i}^{max}x_{i}^{on/off}(t) \geqslant P_{res}(t), \forall t\in T,
\end{equation}
\end{linenomath*}
\begin{linenomath*}
\begin{equation} 
\label{eq:Dispatch5}
	\sum_{\tau = t}^{t+T_i^{up}-1} x_{i}^{on/off}(\tau) \geqslant T_i^{up}x_{i}^{up}(t), \forall i\in \Phi, \forall t\in T,
\end{equation}
\end{linenomath*}
\begin{linenomath*}
\begin{equation} 
\label{eq:Dispatch6}
	\sum_{\tau = t}^{t+T_i^{down}-1} [1-x_{i}^{on/off}(\tau)] \geqslant T_i^{down}x_{i}^{down}(t), \forall i\in \Phi, \forall t\in T,
\end{equation}
\end{linenomath*}
\begin{linenomath*}
\begin{equation} 
\label{eq:Dispatch7}
	x_{i}^{up}(t) + x_{i}^{down}(t) \leqslant 1, \forall i\in \Phi, \forall t\in T,
\end{equation}
\end{linenomath*}
\begin{linenomath*}
\begin{equation} 
\label{eq:Dispatch8}
	x_{i}^{up}(t) - x_{i}^{down}(t) = x_{i}^{on/off}(t) - x_{i}^{on/off}(t-1), \forall i\in \Phi, \forall t\in T,
\end{equation}
\end{linenomath*}
\begin{linenomath*}
\begin{equation} 
\label{eq:Dispatch9}
	x_{i}^{power}(t) \geqslant 0, \forall i\in \Phi, \forall t\in T,
\end{equation}
\end{linenomath*}
\begin{linenomath*}
\begin{equation} 
\label{eq:Dispatch10}
	x_{i}^{up}(t),x_{i}^{down}(t),x_{i}^{on/off}(t) \in \{0,1\}, \forall i\in \Phi, \forall t\in T,
\end{equation}
\end{linenomath*}
where $\Phi$ is the set of coal-fired power units and $T=\{1,2,...,48\}$ represents the time slots among a day's 48 half-hours. $X=\{x_{i}^{power}(t),x_{i}^{up}(t),x_{i}^{down}(t),x_{i}^{on/off}(t)\}$ are decision variables. $x_{i}^{power}(t)$ indicates the scheduled power generation of coal-fired power unit $i$ at time slot $t$. $x_{i}^{up}(t)$ and $x_{i}^{down}(t)$ denote the startup and shutdown operations of the coal-fired power unit $i$ at time slot $t$. $x_{i}^{on/off}(t)$ denotes the on/off states of the coal-fired unit $i$ at time slot $t$. The objective \eqref{eq:Dispatch1} is to minimize the total economic cost, including the power generation cost $c_i^{power}x_{i}^{power}(t)$, the startup cost $c_i^{up}x_{i}^{up}(t)$ and shutdown cost $c_i^{down}x_{i}^{down}(t)$. \eqref{eq:Dispatch2} represents the equilibrium between power generation and consumption. $P_{outside}(t)$ denotes the input power from outside regions. $P_{load}(t)$ denotes the local power load at time slot $t$. $r_{loss}$ denotes the average loss rate of power transmission. \eqref{eq:Dispatch3} specifies the upper and lower bounds of power generation. \eqref{eq:Dispatch4} regulates that the total capacity of working units must exceed the spinning reserve requirement $P_{res}(t)$. \eqref{eq:Dispatch5} and~\eqref{eq:Dispatch6} indicate that each unit must maintain a working state for a minimum of half an hour. $T_i^{up}$ and $T_i^{down}$ denote the minimum on and off half-hours of power unit $i\in \Phi$. \eqref{eq:Dispatch7} and~\eqref{eq:Dispatch8} show the connections between startup/shutdown operations and each power unit's on/off state. In \eqref{eq:Dispatch10}, $x_{i}^{up}(t)=1$ indicates a startup operation. $x_{i}^{down}(t)=1$ indicates a shutdown operation. $x_{i}^{on/off}(t)=1$ indicates that the power unit $i$ is on. 

{In Nanchang, the source of electricity has two parts. One is from the local power plants distributed in Nanchang ($x_{i}^{power}, i\in \Phi$), and the other is from the input power from outside regions ($P_{outside}$). The local power plants in Nanchang are controllable in the electricity dispatch model, which support the energy consumption of mobile networks. According to the government report~\cite{Nanchang:2021}, Nanchang has three local power plants: the Nanchang plant, the Xinchang plant, and the Hongping plant.}

The annual development report of China's power industry~\cite{Report:2021} shows that the thermal coal consumption rate for generation with different capacities is,
\begin{linenomath*}
\begin{equation} 
\label{eq:ElectricData3}
	c_{i}^{coal} = 
	\begin{cases}
		0.3007\; \text{t/MWh},  &  P_{i}^{max}>300\; \text{MW} \\
		0.3357\; \text{t/MWh},  &  P_{i}^{max}\leqslant 300\; \text{MW}
	\end{cases}.
\end{equation}
\end{linenomath*}
By optimizing \eqref{eq:Dispatch1}, the optimal scheduling strategies for coal-fired power units can be obtained. Let $x_i^{power*}(t)$ denote the optimal solution. The total thermal coal consumption within one day, denoted by $Q^{Coal}$, can be calculated as follows:
\begin{linenomath*}
\begin{equation}
\label{eq:coal}
    Q^{Coal} = \sum_{i \in \Phi}\sum_{t \in T}\frac{1}{2}c_{i}^{coal}x_i^{power*}(t).
\end{equation}
\end{linenomath*}
The coefficient $1/2$ refers to half an hour because the time slot in our case is in half an hour.

\noindent
\textbf{Carbon Emission Estimation for Mobile Networks.} To examine the carbon emissions from mobile networks, the total power load $P_{load}(t)$ is divided into two parts:
\begin{linenomath*}
\begin{equation} 
\label{eq:ElectricData4}
	P_{load}(t) = P_{orig}(t) + P_{BS}(t), \forall t\in T,
\end{equation}
\end{linenomath*}
where $P_{orig}(t)$ is the residential and industrial power load excluding base station power load and $P_{BS}(t)$ is the power dispatched to base stations to meet their operational needs. The difference in carbon emissions with and without $P_{BS}(t)$ is used to characterize the carbon emission caused by mobile networks. Specifically, we first solve \eqref{eq:Dispatch1} with $P_{load}(t) = P_{orig}(t) + P_{BS}(t)$ and obtain the optimal solution of power generation $x_{i}^{power*(1)}(t)$. We next get the optimal solution of power generation $x_{i}^{power*(2)}(t)$ with $P_{load}(t) = P_{orig}(t)$. In terms of \eqref{eq:coal}, the thermal coal consumption of mobile networks within one day, denoted by $Q_{Mobile}^{Coal}$, is computed as follows,
\begin{linenomath*}
\begin{equation} 
\label{eq:ElectricData6}
    Q_{Mobile}^{Coal} = \sum_{i \in \Phi}\sum_{t \in T}\frac{1}{2}c_{i}^{coal}\left[x_i^{power*(1)}(t)- x_i^{power*(2)}(t)\right],
\end{equation}
\end{linenomath*}

{We consider China's thermal coal, which consists of anthracite, bituminous, lignite, etc. Liu et al.~\cite{liu2015reduced} pointed out that China's thermal coal has its unique low heating values ($h^{coal}$ = 20.95 GJ/T) comparing with the global average heating value (29.3 GJ/T) provided by the United Nations. The net carbon content per energy is $\alpha^{coal}$ = 26.59 tC/TJ, and the oxidization rate of thermal coal is set as $O^{coal}$ = 99\%. Thus, the emission factor of China's thermal coal, denoted by $e^{coal}$, is calculated as follows,
\begin{linenomath*}
\begin{equation} 
   \label{eq:ElectricData5}
	e^{coal} = o^{coal} \cdot h^{coal} \cdot \alpha^{coal} \cdot 3.67 = 2.02\;\; \text{tCO}_2/\text{t}.
\end{equation}
\end{linenomath*}}
The carbon emissions from mobile networks, denoted by $Q_{Mobile}^{CO_2}$, is computed as follows:
\begin{linenomath*}
\begin{equation}
    Q_{Mobile}^{CO_2} = e^{coal}Q_{Mobile}^{Coal}.
\end{equation}
\end{linenomath*}

In practice, the $P_{orig}(t)$ of Nanchang is approximated through the yearly and daily typical power load curves of Jiangxi province (see Supplementary Note 3 for details). Also, we assumed that solar panels are installed on base stations, which can supply a portion of the base stations' power load. Therefore, the power dispatched to a single base station is expressed as,
\begin{linenomath*}
\begin{equation} 
\label{eq:ElectricData10}
	P_{BS,j}(t) = 
	\begin{cases}
     	\hat{P}_{BS,j}(t)-PV_{BS,j}(t), & \text{if}\;\; \hat{P}_{BS,j}(t)>PV_{BS,j}(t) \\ 
		0,  &  \text{otherwise}
	\end{cases},
\end{equation}
\end{linenomath*}
where $\hat{P}_{BS,j}(t)$ denotes the power load of base station $j$ at time slot $t \in T$, $PV_{BS,j}(t)$ denotes the power generated by the solar panel installed on base station $j$, and ${P}_{BS,j}(t)$ denotes the power dispatched to base station $j$ from power plants. Therefore, the power dispatched to base stations $P_{BS}(t)$ can be expressed as,
\begin{linenomath*}
\begin{equation} \label{ElectricData11}
	P_{BS}(t) = \sum_{j}P_{BS,j}(t).
\end{equation}
\end{linenomath*}

\subsubsection*{M2.2. Carbon Emission Estimation Across Provinces}

In this paper, we generalize the simulation results from Nanchang to all of China's provinces. We first estimate each province's network capacity and traffic through the number of base stations and mobile users. By assuming traffic and energy misalignment factors are comparable across different cities, we then estimate the energy efficiency for each province. We further estimate the energy consumption of mobile networks by dividing network traffic by energy efficiency. Lastly, we convert energy consumption to carbon emissions using the grid emission factors derived from Nanchang. 

{We apply the Monte Carlo method to estimate the capacity of mobile networks in each province. We repeatedly and randomly sample base stations from the Nanchang set according to the number of 4G base stations and 5G base stations in each province. The Monte Carlo simulations are performed 1,000 times for each province.} By adding each sampled base station's capacity and maximum energy consumption, we can obtain the total network capacity $C_p$ and maximum energy consumption $E_{Max, p}$ of mobile networks in province $p$.

We next estimate the network traffic of each province. We assumed that, for a given month, the average traffic per user across regions is comparable. Thus, we determine the average traffic per user in Nanchang.
\begin{linenomath*}
\begin{equation}
    \bar{L} = ({{L}_{4G} + {L}_{5G}})/{N_{{user}}},
\end{equation}
\end{linenomath*}
where $\bar{L}$ denotes the average traffic per user, ${L}_{4G}$ and ${L}_{5G}$ refer to the traffic load of 4G and 5G networks in Nanchang, and $N_{{user}}$ denotes the number of mobile users in Nanchang (Supplementary Table 2). We next estimate provincial network traffic based on the number of mobile users in each province (Supplementary Table 4),
\begin{linenomath*}
\begin{equation}
    {L}_p = \bar{L} \cdot {N_{p,{user}}},
\end{equation}
\end{linenomath*}
where ${L}_p$ and $N_{p,{user}}$ denote the traffic load and the number of mobile users in province $p$, respectively.  

Given the misalignment factor $M_{p}$ of province $p$, the energy efficiency of the mobile network in province $p$, when the network traffic load is $L_p$, can be expressed as,
\begin{linenomath*}
\begin{equation} 
 {\eta}^{Energy}_p(L_p) =  \frac{C_p/E_{Max,p}} {(1 + M_{p}/\tilde{L}_p)},
\end{equation}
\end{linenomath*}
where $\tilde{L}_p = L_p / C_p$ denotes the normalized traffic load. {We estimate the misalignment factor $M_{p}$ of province $p$ based on its normalized network traffic load and the energy-saving
method it uses (see Supplementary Note 5 for details)}. The energy consumption of the mobile network in province $p$, denoted by $E_p$, can be estimated as,
\begin{linenomath*}
\begin{equation}
    E_p = E_{Max,p} (M_p + \tilde{L}_p).
\end{equation}
\end{linenomath*}
{According to Nanchang's results, the average grid emission factor is $ \gamma^{CO_2} = 0.68~\text{tCO}_2/\text{MWh}$, which has no significant difference before and after launching the 5G network (Supplementary \figurename 5)}. Thus, the carbon emissions of mobile networks in province $p$, denoted by $Q_{Mobile, p}^{CO_2}$, is computed as follows:
\begin{linenomath*}
\begin{equation}
    Q_{Mobile, p}^{CO_2} = E_p \cdot \gamma^{CO_2}.
\end{equation}
\end{linenomath*}

\subsection*{M3. Energy-Saving Method in Mobile Networks}

DeepEnergy divides all cells into multiple grids and then trains reinforcement learning (RL) agents to decide whether each cell should enter sleep mode at each time step (see Supplementary Note 4 for details). A cell refers to a carrier on a sector of a base station. Typically, a base station supports multiple sectors. Thus, a base station has multiple cells. 
The RL agents need to control the status of cells based on the current state of the mobile network to reduce the total amount of energy consumption. This includes the energy consumed by RRUs in cells and the energy consumed by BBUs and cooling devices in base stations.
We model cell control as a multi-agent cooperation problem: each cell $c_n$ is managed by an agent ${agent}_n$, and these agents cooperate to minimize energy consumption.

The observation of each agent ${agent}_n$ includes the feature vectors of other cells in the same base station or the same grid. The feature vector of cell $c_n$ consists of time $t$, the traffic loads of the grid $g_m$ that the cell $c_n$ belongs to in the last four time steps, $(L_{t-4}^m, L_{t-3}^m, L_{t-2}^m, L_{t-1}^m)$, and the device parameters of the cells in $\mathcal{N}_m^g \cup \mathcal{N}_k^b$. $\mathcal{N}_m^g$ denotes the set of cells in the grid $g_m$ and $\mathcal{N}_k^b$ denotes the set of cells of the base station $BS_k$ that the cell $c_n$ belongs to.

To encourage agents to work together, we set the reward for agent $c_n$ to $r_n=-\sum_{n'\in{\mathcal{N}_m^g}} P^{RRU}_{n'} - P^{BBU}_{k}-P^{cooling}_{k}$ (see Supplementary Note 4 for details).
The first term indicates the energy of RRUs in cells that belong to the same grid as cell $c_n$, so it could encourage the agent to cooperate with agents from these cells $\{c_{n'}\}_{n'\in{\mathcal{N}_m^g}}$.
The second term indicates the energy of BBUs and cooling devices in the base station that cell $c_n$ belongs to, which may encourage the agent to cooperate with the agents from the same base station.

We consider training a policy network to output the action based on the current state to solve this multi-agent collaboration problem.
However, the large number of agents will be a major challenge. On the one hand, the large-scale agents lead to massive state space. When an agent makes the decision, it should efficiently integrate the state of other associated agents. 
For this problem, we construct two graphs based on the intra-grid and intra-BS relationships and then use the graph neural network to integrate the information of the associated nodes (i.e., agents) in the graphs.
On the other hand, large-scale agents also lead to complex interactions between agents, making the expected rewards of different actions difficult to estimate. Thus, we introduce the idea of Mean-Field MARL~(MF-MARL~\cite{yang2018mean}) to integrate the effects of other agents' decisions on the target agent and design two masks to decrease the difficulty of action value estimation. The first mask denotes whether the grid's traffic demand could be met if the agent sets the respective cell to sleep.
The second mask denotes whether all other cells under the same base station could sleep if the cell is set to sleep mode.

These masks are derived from the actions of other agents and can only be used as training input. Therefore, each agent has two action-value networks, $q^2(\mathbf{o}_n,a,\mathbf{m}_n;\theta_q^2)$ for training and $q^1(\mathbf{o}_n,a;\theta_q^1)$ for inference, where $\mathbf{o}$ denotes the observation, $a$ denotes the action, $\mathbf{m}$ denotes masks, and $\theta$ denotes the parameters of action-value networks. 
In the execution stage, the action-value network without masks as input, $q^1(\mathbf{o}_n, a;\theta_q^1)$, predicts the rewards associated with different actions, setting the cell to sleep or not, based on the observation.
Using $\epsilon$-greedy method, the agent determines the status of the cell by sampling the action based on predicted rewards.
After the statuses of all cells determined by the agents are executed, the rewards for all agents are calculated. All agents' observations, actions, and rewards are appended to the replay buffer.
In the training stage, the action-value network with masks as input, $q^2(\mathbf{o}_n, a, \mathbf{m}_n;\theta_q^2)$, is trained with the replay buffer to predict the rewards of different actions based on the observation and masks.
The network $q^2$ is optimized through the following loss function:
\begin{linenomath*}
\begin{equation}\label{equ:rl_l2}
    L_2 = \frac{1}{2} \big[q^2(\mathbf{o}_n,a,\mathbf{m}_n;\theta_q^2)-r_n\big]^2,
\end{equation}
\end{linenomath*}
where the masks $\mathbf{m}_n$ is obtained based on the actions recorded in replay buffer.
The action-value network, $q^1(\mathbf{o}_n,a;\theta_q^1)$, is then optimized by imitating the second one, $q^2(\mathbf{o}_n,a,\mathbf{m}_n;\theta_q^2)$,
\begin{linenomath*}
\begin{equation}\label{equ:rl_l1}
    L_1 = \frac{1}{2} \sum_a \big[q^1(\mathbf{o}_n,a;\theta_q^1)-q^2(\mathbf{o}_n,a,\mathbf{m}_n;\theta_q^2)\big]^2,
\end{equation}
\end{linenomath*}
where the mask $\mathbf{m}_n$ are calculated based on the re-sampled actions, which is sampled as the process in the execution stage.
The re-sampling process is equivalent to the sampling process for the distribution of the mask vector. Thus, the first network $q^1$ would learn to predict the rewards of various actions under the sampled distribution of the mask vector. The framework of DeepEnergy is shown in Supplementary ~\figurename~19. The algorithm is shown in Supplementary Algorithm~1. {Notably, after determining the working state of cells, we still need to re-distribute intra-grid network traffic among cells. The cells in each grid are equivalent in terms of service capability and thus intra-grid network traffic can be flexibly transferred between cells within the grid. Specifically, given the working state of cells, we will then re-distribute the intra-grid traffic of each active cell proportionally according to its capacity; namely, the cell with a higher capacity will carry more traffic.}

\subsection*{M4. Solar Photovoltaic Modeling and Simulation}

We use PVWatts~\cite{freeman2018system, waldman2019solar}, a photovoltaic system simulator developed by NREL (The National Renewable Energy Laboratory), to simulate the potential electric generation of a PV system. We locate PV panels using the coordinates of base stations and use the typical meteorological year data for simulation. Defaults set all other parameters as PVWatts provides the most up-to-date standards for PV installations (see Supplementary Table 5 for details). 

{The cost structure of a PV system is divided into the initial investment cost, the operation cost, and maintenance cost~\cite{Yan2019}. The initial investment cost includes PV modules, supporting structures, inverters, wiring, junction boxes, engineering costs (design, transportation, and installation), and insurance. The annual operating costs are equivalent to 1\% of system investment. The maintenance cost includes inverter replacement in the tenth year~\cite{kaabeche2011techno}. We assume that the lifetime of the PV system is 20 years. The specific costs are listed in Supplementary Table 17, which are compiled from the China Photovoltaic Industry Association (China PV industry development roadmap; \href{http://www.chinapv.org.cn/road_map.html}{http://www.chinapv.org.cn/road$\_$map.html}). The cost data is real-world data collected from the photovoltaic industry of China by the China Photovoltaic Industry Association, and under the guidance of Ministry of Industry and Information Technology.}

\section*{Data Availability}

The number of base stations and mobile users in each province are listed in Supplementary Tables 3 and 4. The network traffic data and the number of mobile users in Nanchang are listed in Supplementary Table 2. Source data are provided with this paper.

\section*{Code Availability}

The code used in this study can be downloaded from \href{https://github.com/Tong89/Sustainability\_5G}{https://github.com/Tong89/Sustainability\_5G}.

\bibliography{Ref}

\begin{thebibliography}{10}
\urlstyle{rm}
\expandafter\ifx\csname url\endcsname\relax
  \def\url#1{\texttt{#1}}\fi
\expandafter\ifx\csname urlprefix\endcsname\relax\def\urlprefix{URL }\fi
\expandafter\ifx\csname doiprefix\endcsname\relax\def\doiprefix{DOI: }\fi
\providecommand{\bibinfo}[2]{#2}
\providecommand{\eprint}[2][]{\url{#2}}

\bibitem{manyika2011great}
\bibinfo{author}{Manyika, J.} \& \bibinfo{author}{Roxburgh, C.}
\newblock \bibinfo{journal}{\bibinfo{title}{The great transformer: The impact
  of the internet on economic growth and prosperity}}.
\newblock {\emph{\JournalTitle{McKinsey Global Institute}}}
  \textbf{\bibinfo{volume}{1}} (\bibinfo{year}{2011}).

\bibitem{dang2020should}
\bibinfo{author}{Dang, S.}, \bibinfo{author}{Amin, O.},
  \bibinfo{author}{Shihada, B.} \& \bibinfo{author}{Alouini, M.-S.}
\newblock \bibinfo{journal}{\bibinfo{title}{What should {6G} be?}}
\newblock {\emph{\JournalTitle{Nature Electronics}}}
  \textbf{\bibinfo{volume}{3}}, \bibinfo{pages}{20--29} (\bibinfo{year}{2020}).

\bibitem{mwangama2020can}
\bibinfo{author}{Mwangama, J.}, \bibinfo{author}{Malila, B.},
  \bibinfo{author}{Douglas, T.} \& \bibinfo{author}{Rangaka, M.}
\newblock \bibinfo{journal}{\bibinfo{title}{What can {5G} do for healthcare in
  africa?}}
\newblock {\emph{\JournalTitle{Nature Electronics}}}
  \textbf{\bibinfo{volume}{3}}, \bibinfo{pages}{7--9} (\bibinfo{year}{2020}).

\bibitem{gohar2021role}
\bibinfo{author}{Gohar, A.} \& \bibinfo{author}{Nencioni, G.}
\newblock \bibinfo{journal}{\bibinfo{title}{The role of {5G} technologies in a
  smart city: The case for intelligent transportation system}}.
\newblock {\emph{\JournalTitle{Sustainability}}} \textbf{\bibinfo{volume}{13}},
  \bibinfo{pages}{5188} (\bibinfo{year}{2021}).

\bibitem{dai2019human}
\bibinfo{author}{Dai, C.}, \bibinfo{author}{Liu, X.}, \bibinfo{author}{Lai,
  J.}, \bibinfo{author}{Li, P.} \& \bibinfo{author}{Chao, H.-C.}
\newblock \bibinfo{journal}{\bibinfo{title}{Human behavior deep recognition
  architecture for smart city applications in the {5G} environment}}.
\newblock {\emph{\JournalTitle{IEEE Network}}} \textbf{\bibinfo{volume}{33}},
  \bibinfo{pages}{206--211} (\bibinfo{year}{2019}).

\bibitem{taleb2019orchestrating}
\bibinfo{author}{Taleb, T.}, \bibinfo{author}{Afolabi, I.} \&
  \bibinfo{author}{Bagaa, M.}
\newblock \bibinfo{journal}{\bibinfo{title}{Orchestrating {5G} network slices
  to support industrial internet and to shape next-generation smart
  factories}}.
\newblock {\emph{\JournalTitle{IEEE Network}}} \textbf{\bibinfo{volume}{33}},
  \bibinfo{pages}{146--154} (\bibinfo{year}{2019}).

\bibitem{GSA2022}
\bibinfo{author}{{Global mobile suppliers association}}.
\newblock \bibinfo{title}{{5G}-market snapshot june 2022}.
\newblock
  \bibinfo{howpublished}{\url{https://gsacom.com/paper/5G-market-snapshot-june-2022/}}
  (\bibinfo{year}{2022}).

\bibitem{Juan2022}
\bibinfo{author}{Tomás, J.~P.}
\newblock \bibinfo{title}{China ends august with 2.1 million {5G} base
  stations: Report}.
\newblock
  \bibinfo{howpublished}{\url{https://www.rcrwireless.com/20220923/5G/china-ends-august-2-million-5G-base-stations-report}}
  (\bibinfo{year}{2022}).

\bibitem{al2022massive}
\bibinfo{author}{Al-Bawri, S.~S.}, \bibinfo{author}{Islam, M.~T.},
  \bibinfo{author}{Islam, M.~S.}, \bibinfo{author}{Singh, M.~J.} \&
  \bibinfo{author}{Alsaif, H.}
\newblock \bibinfo{journal}{\bibinfo{title}{Massive metamaterial system-loaded
  mimo antenna array for {5G} base stations}}.
\newblock {\emph{\JournalTitle{Scientific Reports}}}
  \textbf{\bibinfo{volume}{12}}, \bibinfo{pages}{1--16} (\bibinfo{year}{2022}).

\bibitem{hecht2016bandwidth}
\bibinfo{author}{Hecht, J.} \emph{et~al.}
\newblock \bibinfo{journal}{\bibinfo{title}{The bandwidth bottleneck}}.
\newblock {\emph{\JournalTitle{Nature}}} \textbf{\bibinfo{volume}{536}},
  \bibinfo{pages}{139--142} (\bibinfo{year}{2016}).

\bibitem{han2020energy}
\bibinfo{author}{I, C.-L.}, \bibinfo{author}{Han, S.} \& \bibinfo{author}{Bian,
  S.}
\newblock \bibinfo{journal}{\bibinfo{title}{Energy-efficient {5G} for a greener
  future}}.
\newblock {\emph{\JournalTitle{Nature Electronics}}}
  \textbf{\bibinfo{volume}{3}}, \bibinfo{pages}{182--184}
  (\bibinfo{year}{2020}).

\bibitem{ding2022carbon}
\bibinfo{author}{Ding, Y.} \emph{et~al.}
\newblock \bibinfo{journal}{\bibinfo{title}{Carbon emissions and mitigation
  potentials of {5G} base station in china}}.
\newblock {\emph{\JournalTitle{Resources, Conservation and Recycling}}}
  \textbf{\bibinfo{volume}{182}}, \bibinfo{pages}{106339}
  (\bibinfo{year}{2022}).

\bibitem{ilieva2018social}
\bibinfo{author}{Ilieva, R.~T.} \& \bibinfo{author}{McPhearson, T.}
\newblock \bibinfo{journal}{\bibinfo{title}{Social-media data for urban
  sustainability}}.
\newblock {\emph{\JournalTitle{Nature Sustainability}}}
  \textbf{\bibinfo{volume}{1}}, \bibinfo{pages}{553--565}
  (\bibinfo{year}{2018}).

\bibitem{yang2016interference}
\bibinfo{author}{Yang, C.}, \bibinfo{author}{Li, J.}, \bibinfo{author}{Ni, Q.},
  \bibinfo{author}{Anpalagan, A.} \& \bibinfo{author}{Guizani, M.}
\newblock \bibinfo{journal}{\bibinfo{title}{Interference-aware energy
  efficiency maximization in {5G} ultra-dense networks}}.
\newblock {\emph{\JournalTitle{IEEE Transactions on Communications}}}
  \textbf{\bibinfo{volume}{65}}, \bibinfo{pages}{728--739}
  (\bibinfo{year}{2016}).

\bibitem{Francestatista}
\bibinfo{author}{Tiseo, I.}
\newblock \bibinfo{title}{{Power Sector Emissions in France from 2000 to
  2021}}.
\newblock
  \bibinfo{howpublished}{\url{https://www.statista.com/statistics/1290541/power-sector-carbon-emissions-france/}}
  (\bibinfo{year}{2022}).

\bibitem{birdsey1992carbon}
\bibinfo{author}{Birdsey, R.~A.}
\newblock \emph{\bibinfo{title}{Carbon storage and accumulation in United
  States forest ecosystems}}, vol.~\bibinfo{volume}{59} (\bibinfo{publisher}{US
  Department of Agriculture, Forest Service}, \bibinfo{year}{1992}).

\bibitem{peng2011traffic}
\bibinfo{author}{Peng, C.}, \bibinfo{author}{Lee, S.-B.}, \bibinfo{author}{Lu,
  S.}, \bibinfo{author}{Luo, H.} \& \bibinfo{author}{Li, H.}
\newblock \bibinfo{title}{Traffic-driven power saving in operational {3G}
  cellular networks}.
\newblock In \emph{\bibinfo{booktitle}{Proceedings of the 17th annual
  international conference on Mobile computing and networking}},
  \bibinfo{pages}{121--132} (\bibinfo{year}{2011}).

\bibitem{rostami2019pre}
\bibinfo{author}{Rostami, S.}, \bibinfo{author}{Heiska, K.},
  \bibinfo{author}{Puchko, O.}, \bibinfo{author}{Leppanen, K.} \&
  \bibinfo{author}{Valkama, M.}
\newblock \bibinfo{journal}{\bibinfo{title}{Pre-grant signaling for
  energy-efficient 5g and beyond mobile devices: Method and analysis}}.
\newblock {\emph{\JournalTitle{IEEE Transactions on Green Communications and
  Networking}}} \textbf{\bibinfo{volume}{3}}, \bibinfo{pages}{418--432}
  (\bibinfo{year}{2019}).

\bibitem{wu2015energy}
\bibinfo{author}{Wu, J.}, \bibinfo{author}{Zhang, Y.},
  \bibinfo{author}{Zukerman, M.} \& \bibinfo{author}{Yung, E. K.-N.}
\newblock \bibinfo{journal}{\bibinfo{title}{Energy-efficient base-stations
  sleep-mode techniques in green cellular networks: A survey}}.
\newblock {\emph{\JournalTitle{IEEE communications surveys \& tutorials}}}
  \textbf{\bibinfo{volume}{17}}, \bibinfo{pages}{803--826}
  (\bibinfo{year}{2015}).

\bibitem{yu2014dual}
\bibinfo{author}{Yu, G.}, \bibinfo{author}{Chen, Q.} \& \bibinfo{author}{Yin,
  R.}
\newblock \bibinfo{journal}{\bibinfo{title}{Dual-threshold sleep mode control
  scheme for small cells}}.
\newblock {\emph{\JournalTitle{IET communications}}}
  \textbf{\bibinfo{volume}{8}}, \bibinfo{pages}{2008--2016}
  (\bibinfo{year}{2014}).

\bibitem{chamola2016solar}
\bibinfo{author}{Chamola, V.} \& \bibinfo{author}{Sikdar, B.}
\newblock \bibinfo{journal}{\bibinfo{title}{Solar powered cellular base
  stations: current scenario, issues and proposed solutions}}.
\newblock {\emph{\JournalTitle{IEEE Communications magazine}}}
  \textbf{\bibinfo{volume}{54}}, \bibinfo{pages}{108--114}
  (\bibinfo{year}{2016}).

\bibitem{freeman2018system}
\bibinfo{author}{Freeman, J.~M.} \emph{et~al.}
\newblock \bibinfo{title}{System advisor model (sam) general description
  (version 2017.9. 5)}.
\newblock \bibinfo{type}{Tech. Rep.}, \bibinfo{institution}{National Renewable
  Energy Lab.(NREL), Golden, CO (United States)} (\bibinfo{year}{2018}).

\bibitem{waldman2019solar}
\bibinfo{author}{Waldman, J.}, \bibinfo{author}{Sharma, S.},
  \bibinfo{author}{Afshari, S.} \& \bibinfo{author}{Fekete, B.}
\newblock \bibinfo{journal}{\bibinfo{title}{Solar-power replacement as a
  solution for hydropower foregone in us dam removals}}.
\newblock {\emph{\JournalTitle{Nature Sustainability}}}
  \textbf{\bibinfo{volume}{2}}, \bibinfo{pages}{872--878}
  (\bibinfo{year}{2019}).

\bibitem{peng2018managing}
\bibinfo{author}{Peng, W.} \emph{et~al.}
\newblock \bibinfo{journal}{\bibinfo{title}{Managing china’s coal power
  plants to address multiple environmental objectives}}.
\newblock {\emph{\JournalTitle{Nature Sustainability}}}
  \textbf{\bibinfo{volume}{1}}, \bibinfo{pages}{693--701}
  (\bibinfo{year}{2018}).

\bibitem{Wang:20}
\bibinfo{author}{Wang, J.} \emph{et~al.}
\newblock \bibinfo{journal}{\bibinfo{title}{Exploring the trade-offs between
  electric heating policy and carbon mitigation in china}}.
\newblock {\emph{\JournalTitle{Nature Communications}}}
  \textbf{\bibinfo{volume}{11}}, \bibinfo{pages}{1--11},
  \doiprefix\url{10.1038/s41467-020-19854-y} (\bibinfo{year}{2020}).

\bibitem{bogdanov2019radical}
\bibinfo{author}{Bogdanov, D.} \emph{et~al.}
\newblock \bibinfo{journal}{\bibinfo{title}{Radical transformation pathway
  towards sustainable electricity via evolutionary steps}}.
\newblock {\emph{\JournalTitle{Nature communications}}}
  \textbf{\bibinfo{volume}{10}}, \bibinfo{pages}{1--16} (\bibinfo{year}{2019}).

\bibitem{miitgov}
\bibinfo{author}{{People's Republic of China's Ministry of Industry and
  Information Technology}}.
\newblock \bibinfo{title}{Statistics in communication industry}.
\newblock
  \bibinfo{howpublished}{\url{https://www.miit.gov.cn/gxsj/tjfx/txy/index.html}}
  (\bibinfo{year}{2022}).

\bibitem{arnold2010power}
\bibinfo{author}{Arnold, O.}, \bibinfo{author}{Richter, F.},
  \bibinfo{author}{Fettweis, G.} \& \bibinfo{author}{Blume, O.}
\newblock \bibinfo{title}{Power consumption modeling of different base station
  types in heterogeneous cellular networks}.
\newblock In \emph{\bibinfo{booktitle}{2010 Future Network \& Mobile Summit}},
  \bibinfo{pages}{1--8} (\bibinfo{organization}{IEEE}, \bibinfo{year}{2010}).

\bibitem{CL2014Toward}
\bibinfo{author}{I, C.-L.} \emph{et~al.}
\newblock \bibinfo{journal}{\bibinfo{title}{Toward green and soft: a {5G}
  perspective}}.
\newblock {\emph{\JournalTitle{IEEE Communications Magazine}}}
  \textbf{\bibinfo{volume}{52}}, \bibinfo{pages}{66--73},
  \doiprefix\url{10.1109/MCOM.2014.6736745} (\bibinfo{year}{2014}).

\bibitem{Huang2020Prospect}
\bibinfo{author}{Huang, Y.}, \bibinfo{author}{Xu, X.}, \bibinfo{author}{Li,
  N.}, \bibinfo{author}{Ding, H.} \& \bibinfo{author}{Tang, X.}
\newblock \bibinfo{journal}{\bibinfo{title}{Prospect of {5G} intelligent
  networks}}.
\newblock {\emph{\JournalTitle{IEEE Wireless Communications}}}
  \textbf{\bibinfo{volume}{27}}, \bibinfo{pages}{4--5},
  \doiprefix\url{10.1109/MWC.2020.9170260} (\bibinfo{year}{2020}).

\bibitem{Lopez-Perez2022}
\bibinfo{author}{Lopez-Perez, D.} \emph{et~al.}
\newblock \bibinfo{journal}{\bibinfo{title}{A survey on {5G} radio access
  network energy efficiency: Massive mimo, lean carrier design, sleep modes,
  and machine learning}}.
\newblock {\emph{\JournalTitle{IEEE Communications Surveys \& Tutorials}}}
  \textbf{\bibinfo{volume}{24}}, \bibinfo{pages}{653--697},
  \doiprefix\url{10.1109/COMST.2022.3142532} (\bibinfo{year}{2022}).

\bibitem{crawley2001energyplus}
\bibinfo{author}{Crawley, D.~B.} \emph{et~al.}
\newblock \bibinfo{journal}{\bibinfo{title}{Energyplus: creating a
  new-generation building energy simulation program}}.
\newblock {\emph{\JournalTitle{Energy and buildings}}}
  \textbf{\bibinfo{volume}{33}}, \bibinfo{pages}{319--331}
  (\bibinfo{year}{2001}).

\bibitem{Nanchang:2021}
\bibinfo{author}{Government, N. M.~P.}
\newblock \bibinfo{title}{Notice of the office of the people's government of
  nanchang city on forwarding the 2021 plan of the municipal development and
  reform commission and state grid nanchang power supply company for orderly
  power consumption of nanchang power grid}.
\newblock \bibinfo{type}{Tech. Rep.}, \bibinfo{institution}{Gazette of Nanchang
  Municipal People's Government} (\bibinfo{year}{2021}).
\newblock
  \bibinfo{note}{\url{http://www.nc.gov.cn/nc_xxgk/jsp/zfgb/ncgb_content.jsp?mid=a731211fe1094abfaea8ab348bd5ba8b}}.

\bibitem{Report:2021}
\bibinfo{author}{Council, C.~E.}
\newblock \emph{\bibinfo{title}{Annual Development Report of China's Power
  Industry (2021)}} (\bibinfo{publisher}{China Building Materials Press},
  \bibinfo{year}{2021}), \bibinfo{edition}{first} edn.

\bibitem{liu2015reduced}
\bibinfo{author}{Liu, Z.} \emph{et~al.}
\newblock \bibinfo{journal}{\bibinfo{title}{Reduced carbon emission estimates
  from fossil fuel combustion and cement production in china}}.
\newblock {\emph{\JournalTitle{Nature}}} \textbf{\bibinfo{volume}{524}},
  \bibinfo{pages}{335--338} (\bibinfo{year}{2015}).

\bibitem{yang2018mean}
\bibinfo{author}{Yang, Y.} \emph{et~al.}
\newblock \bibinfo{title}{Mean field multi-agent reinforcement learning}.
\newblock In \emph{\bibinfo{booktitle}{International conference on machine
  learning}}, \bibinfo{pages}{5571--5580} (\bibinfo{organization}{PMLR},
  \bibinfo{year}{2018}).

\bibitem{Yan2019}
\bibinfo{author}{Yan, J.}, \bibinfo{author}{Yang, Y.},
  \bibinfo{author}{Elia~Campana, P.} \& \bibinfo{author}{He, J.}
\newblock \bibinfo{journal}{\bibinfo{title}{City-level analysis of subsidy-free
  solar photovoltaic electricity price, profits and grid parity in china}}.
\newblock {\emph{\JournalTitle{Nature Energy}}} \textbf{\bibinfo{volume}{4}},
  \bibinfo{pages}{709--717}, \doiprefix\url{10.1038/s41560-019-0441-z}
  (\bibinfo{year}{2019}).

\bibitem{kaabeche2011techno}
\bibinfo{author}{Kaabeche, A.}, \bibinfo{author}{Belhamel, M.} \&
  \bibinfo{author}{Ibtiouen, R.}
\newblock \bibinfo{journal}{\bibinfo{title}{Techno-economic valuation and
  optimization of integrated photovoltaic/wind energy conversion system}}.
\newblock {\emph{\JournalTitle{Solar energy}}} \textbf{\bibinfo{volume}{85}},
  \bibinfo{pages}{2407--2420} (\bibinfo{year}{2011}).

\end{thebibliography}

\section*{Author Contributions Statement}

Tong Li, Depeng Jin, Yong Li and Tao Jiang conceived and designed the study. Li Yu and Yan Zhou collected and provided the data. Tong Li, Yibo Ma, Tong Duan, and Wenzhen Huang carried out the simulations and analyses. All authors contributed to the discussions on the method and the writing of this article.

\section*{Competing Interests}

The authors declare no competing interests.

\newpage
\clearpage{\thispagestyle{empty}\cleardoublepage}
\includepdf[pages=-]{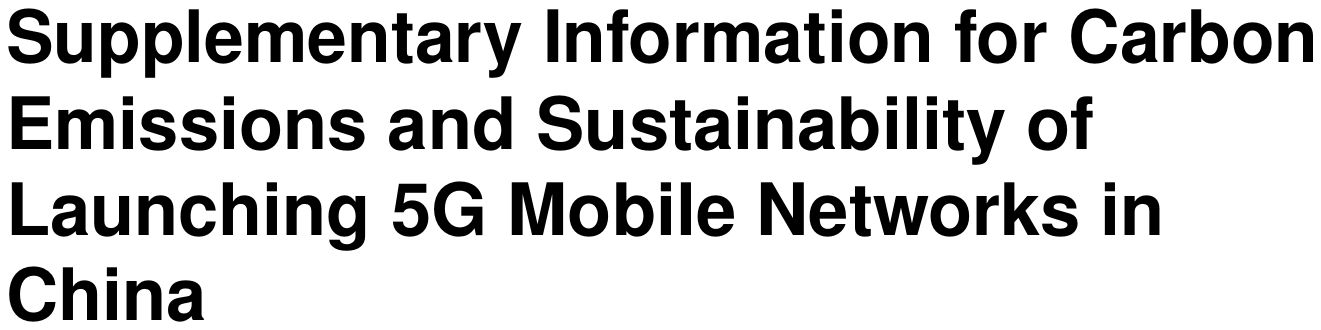}
\end{document}